\documentclass{article}
\usepackage{ieee_tns}
\usepackage[dvips]{epsfig}
\usepackage{cite}

\begin{document}

\hyphenation{ce-ren-kov di-scri-mi-na-tor}

\title{The DISTO Data Acquisition System at SATURNE}

\author{
     F. Balestra$^{d}$, Y. Bedfer$^{c}$, R. Bertini$^{c,d}$, L.C. Bland$^{b}$,
     A. Brenschede$^{h}$, F. Brochard$^{c}$, M.P. Bussa$^{d}$, 
     S. Choi$^{b}$, M. Debowski$^{e}$, M. Dzemidzic$^{b}$, I.V. 
Falomkin$^{a}$,
     J.Cl. Faivre$^{c}$, L. Fava$^{d}$, L. Ferrero$^{d}$, J. Foryciarz$^{g}$, 
     V. Frolov$^{a}$, R. Garfagnini$^{d}$, D. Gill$^{i}$, A. Grasso$^{d}$, 
     E. Grosse$^{e}$, W.W. Jacobs$^{b}$, W. K\"{u}hn$^{h}$,
     A. Maggiora$^{d}$, M. Maggiora$^{d}$, A. Manara$^{b,d}$, 
     D. Panzieri$^{d}$, H. Pfaff$^{h}$, G. Piragino$^{d}$, 
     G.B. Pontecorvo$^{a}$, A. Popov$^{a}$, J. Ritman$^{h}$, 
     P. Salabura$^{f}$, V. Tchalyshev$^{a}$, F. Tosello$^{d}$, 
     S.E. Vigdor$^{b}$, G. Zosi$^{d}$   
      \\
      $^{a}$ JINR - Dubna \\
      $^{b}$ IUCF - Indiana \\
      $^{c}$ LNS - CEN Saclay \\
      $^{d}$ Dipartimento di Fisica``A. Avogadro'' 
                           and INFN - Torino \\
      $^{e}$ GSI - Darmstadt \\
      $^{f}$ Institute of Physics - Krakov \\
      $^{g}$ Jagellonian University - Krakow \\
      $^{h}$ II Physikalisches Institut - Giessen \\
      $^{i}$ TRIUMF - Vancouver 
  }

\maketitle

\begin{abstract}
  The DISTO collaboration has built a large-acceptance magnetic spectrometer 
designed to provide broad kinematic coverage of multi-particle final states 
produced in $pp$ scattering.  The spectrometer has been installed in the 
polarized proton beam of the Saturne accelerator in Saclay to study 
polarization observables in the $\vec{p} p \rightarrow p K^{+} \vec{Y}$ ($Y = 
\Lambda, \Sigma^{0}$ or $Y^{*}$) reaction and vector meson production ($\phi, 
\omega$ and $\rho$) in $pp$ collisions.  The data acquisition system is based 
on a VME 68030 CPU running the OS/9 operating system, housed in a single VME 
crate together with the CAMAC interface, the triple port ECL memories, and 
four RISC R3000 CPU.  The digitization of signals from the detectors is made 
by PCOS III and FERA front-end electronics.  Data of several events belonging 
to a single Saturne extraction are stored in VME triple-port ECL memories 
using a hardwired fast sequencer.  The buffer, optionally filtered by the RISC 
R3000 CPU, is recorded on a DLT cassette by DAQ CPU using the on-board SCSI 
interface during the acceleration cycle.  Two UNIX workstations are connected 
to the VME CPUs through a fast parallel bus and the Local Area Network. 
They analyze a subset of events for on-line monitoring. 
The data acquisition system is able to read and record 3500 ev/burst in the 
present configuration with a dead time of 15\%.
\end{abstract}


\section{Introduction}

\begin{figure}
   \centering\epsfig{figure=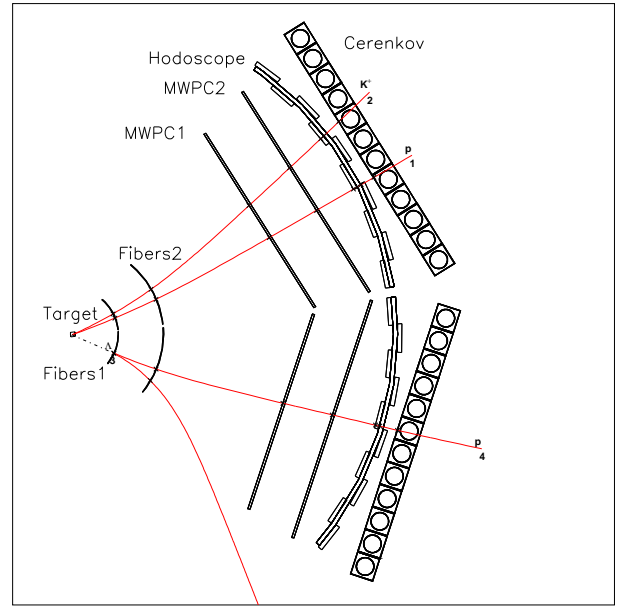,width=.9\linewidth}
\caption{\em Layout of the DISTO experimental setup, shown in plan view 
	 with a simulated $\Lambda$ event.}
\label{fig_expapp}
\end{figure}

The DISTO collaboration has constructed a large-acceptance magnetic 
spectrometer to provide broad kinematic coverage of multi-particle final 
states involving charged particles produced in $pp$ scattering.  
Examples include the measurement of polarization observables in reactions such
as $\vec{p} p \rightarrow p K^{+} \vec{Y}$, with 
$Y$ representing a hyperon or a hyperon resonance ($\Lambda, \Sigma^{0}$ or 
$Y^{*}$), and the study of vector meson ($\phi, \omega$ and $\rho$) production 
in $\vec{p} p$ 
interactions~\cite{arvieux1,abegg,vigdor,arvieux2,bertini,maggiora}.
These measurements are now in progress at Laboratoire Nationale Saturne (LNS) 
in Saclay.  In experiment LNS-E213, the high-quality polarized proton beam, 
with kinetic energies ranging between 1.6 and 2.9 GeV, hits a liquid hydrogen 
target positioned in the center of a magnet that provides a strong magnetic 
field with cylindrical symmetry.  
The Saturne accelerator provides, at 2.9 GeV, every 4 s a spill of protons
with a flat top of 0.5 s. At 1.6 GeV the interspill time can be decreased down
to 2 s.
A sketch of the setup is shown in Fig.~\ref{fig_expapp}.  
Superimposed on the layout is a simulated event of the 
reaction $\vec{p} p \rightarrow p K^{+} \vec{\Lambda}$.

\vspace{2mm}
\vspace{-1mm}

The detectors are arranged in two arms on both sides of the beam pipe (not
shown in the figure). 
Tracking detectors comprise two pairs of cylindrical 
scintillating fiber (SF) chambers 
(two stereo layers, \textit{u-v} planes, and one 
horizontal \textit{y} plane) and two pairs of \textit{u-v-y} planar Multi
Wire Proportional Chambers (MWPC).
The 2492 fibers and 3202 wires of the tracking detectors are equipped with 
discriminator/latch electronics (PCOS III). 
A scintillation counter hodoscope and a
plane of vertically segmented Cerenkov counters
follow the tracking detectors. 
The hodoscope and the fiber chambers feed the level-1 trigger that makes 
decisions based on the multiplicity of charged prongs and on the topology 
of the events.
The 32 scintillation counters
and the 48 Cerenkov counters are equipped with FERA ADC/TDC to allow
readout of pulse height and timing information for particle identification.

\section{Overview of the DISTO DAQ Architecture}

\begin{figure}
   \centering\epsfig{figure=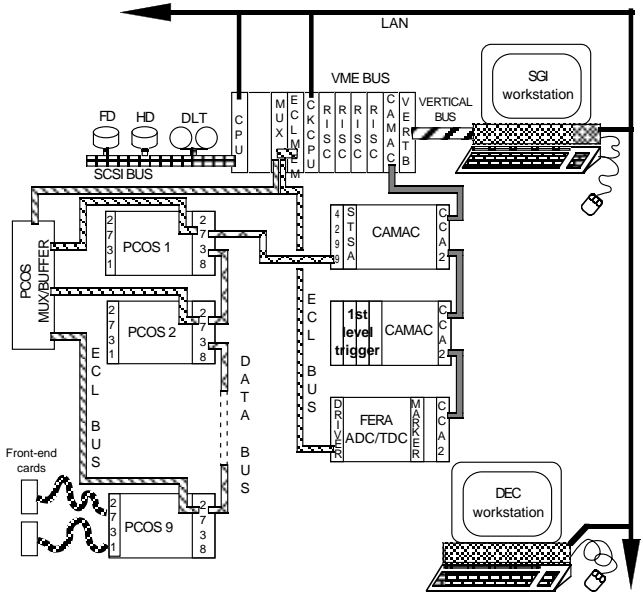,width=\linewidth}
\caption{\em Architecture of the DISTO data acquisition system.}
\label{fig_daq_chain}
\end{figure}

The data acquisition system, shown in Fig.~\ref{fig_daq_chain}, is based on a 
single VME crate, housing the data acquisition (DAQ) CPU, four 
RISC processors and the interface to the 
front-end crates.
The DAQ CPU is connected to a Silicon Graphics (SGI) workstation through a
parallel bus and to a DEC/Alpha workstation
through a VME ELTEC~\cite{eltec} CPU and the Local Area Network (LAN) 
Ethernet line.
The workstations permit on-line monitoring and control of the experiment.
The VME CPUs and the workstations share the same disks using the NFS
protocol.

\vspace{2mm}

The DAQ processor is a commercial VME board equipped with a 68030 CPU,
the CES~\cite{ces} model FIC8232, running the OS/9 operating system.
It performs the following tasks:
\begin{itemize}

  \item set-up of the PCOS III and FERA front-end electronics through a
        standard CAMAC interface;
  \item collection of events from a triple port ECL memory where they have
        been stored by the PCOS and FERA controllers during the Saturne
        extraction time;
  \item dispatching of the events, via the VME bus, to the four RISC R3000
        CPUs;
  \item collection of events for tape recording on DLT cassettes
        through the on-board SCSI interface;
  \item production of ``spy'' events for the monitoring workstations.
\end{itemize}

The on-line programs are written in FORTRAN and C and, as far as the
monitoring is concerned, are based on the PAW~\cite{paw} system developed at 
CERN.
The optional R3000 filter algorithms are written and compiled on the SGI
workstation, taking advantage of the same CPU used in both systems,and
down-loaded through the parallel bus.

\section{DAQ components}

The DISTO on-line data acquisition system architecture 
may be partitioned into several hardware and software components,
discussed separately below.

\subsection{Data Acquisition Hardware Architecture}

The DISTO hardware architecture, shown in Fig.~\ref{fig_daq_chain},
is made up of the following elements 

\subsubsection{Digitization electronics for tracking detectors}

The MWPC  and SF
Multi-anode Photomultiplier's outputs are 
digitized  by the Multiwire Proportional Chamber Operating 
System (PCOS III)~\cite{lecroy}.
The system contains the circuitry to amplify, 
discriminate, delay, latch and encode the signals arriving from 
the tracking detectors.
It includes 16-channel amplifier/discriminator 
cards (N277VG from Nanometrics~\cite{nanom}), 32-channel delay and latch 
modules (Model 2731A), a system controller  (Model 2738)  and a CAMAC DATABUS 
Interface and Buffer (Model 4299).
This system allows very large data rates, by using fast encoding 
to process valid events extremely rapidly and
high-speed transfer of the data at 10 MHz, 10 times the maximum 
rate of CAMAC.

The system uses the facility offered by PCOS III 
to allow computer control of discriminator thresholds and of 
the ripple-thru delays.
The first option is used to measure the efficiency ``plateau'' of the
chambers, the second one allows for an easy tuning of the delay in the
coincidence circuits. 

The 5694 channels are housed in 9 crates. The digitized data containing the
the address of the hit wires are picked-up directly from the ECL output port 
of the system controller.  
The on-board multi-crate interconnection facility, which in principle
allows daisy-chaining of up to 16 crates, was not used due to timing
problem on the ECL ports when more than 5 crates were included in the 
assembly. 
Instead, a new FIFO/MULTIPLEXER module was specially designed to 
record the data 
of the 9 crates in parallel and write them sequentially on a single ECL output 
port. The 9 internal FIFO are able to record the 736 words per crate given 
by a crate completely fired.
The 9 crates are interconnected, also, through the LeCroy DATABUS system. 
It  is employed to send commands and set-up parameters to the PCOS 
system (logical address, threshold, delay and test pattern).


\subsubsection{Digitization electronics for analog signals}

The scintillation and  Cerenkov 
counter's  analog information (charge and time intervals) are 
converted into digital format by the Fast Encoding and readout 
ADC (FERA) and Fast Encoding and Readout TDC (FERET)\cite{lecroy}.
The heart of the system is the module 4300B, a charge 
sensitive 10/11 bit analog to digital converter ( conversion time 
of about $4.8 \mu s$),  the model 4301 FERA Driver and the model 4303 
time-to-charge converter.

The 4301 FERA Driver distributes signals common 
to the system, such as gate, fast clear and handshake signals, via 
the command bus.
It also receives data from the fast data bus, which 
collects data from all model 4300s in the system, and translates 
it for transmission via the on-board ECL-bus (10 Megawords/sec readout).
The last 4300 module of the assembly is a special module able to
write a flag word used as an event separator.
The CAMAC data-way is used for status register and pedestal set-up,
remote testing and control readout modes.

\subsubsection{Event buffering system}

  The data are recorded in the dual daisy-chained triple port memory modules
of the acquisition system via a 30-m long ECL-bus cable for
high speed (10 MHz) readout. 
A special ECL bus multiplexer module was designed to allow fast
switching of PCOS and FERA ECL signals sharing the same VME triple port
memory. 

\subsubsection{Hardware arbitration system}

The PCOS/FERA ECL-bus data multiplexing, the event busy flag signaling 
the digitization and read-out phase in progress, and the CPU read-out
busy signals are made by several commercial and home-made NIM modules. 
The read-out of the buffer stored on the triple-port ECL memory by the data 
acquisition (DAQ) CPU is
triggered at the end of the machine extraction cycle using a CAMAC 
STATUS-A module~\cite{caen}. 
The STATUS-A module is the heart of the synchronization 
between hardware and software, it signals that raw data are
ready to the acquisition task, end-of-readout, selects the acquisition mode 
(single event or multi-event for each readout), and starts, stops, and reset 
the scalers. Another CAMAC crate houses the programmable-threshold 
discriminators and logic matrices to select various level-1 triggers.

\subsubsection{Multiprocessor system}
 
The heart of the hardware architecture is a VME 
crate housing a multiprocessor system, memory and interfaces.
The master CPU is the 68030 equipped with 16 MB 
of dual port RAM and connected by the SCSI bus to a 1GB HD through 
the OS9 operating system; this board carries out three tasks: 
the run-control, the acquisition program and the recording program.

A battery of four RISC CPUs R3000 carry out the optional
run-time analysis and filter task to produce 
good recorded events. The task performed during this
experiment is simply to swap the ECL-memory data.

 A triple port 1 MB memory collects the data (single-event or 
multiple events per burst) from the front-end  PCOS and FERA electronics 
using its ECL input. 
A 8210 CAMAC interface connects  the 68030 CPU  to the CAMAC, 
PCOS and FERA systems.
The vertical bus VIC 8250 is used to insert the Silicon Graphics 
Workstation on the VME-bus, and the ethernet also links it to the 
OS9 68030 CPU.

\subsubsection{Taping system}

The medium used to record the data is ``Digital Linear Tape'' (DLT).
It allows a very high
storage capacity of 20 GBytes with hardware compression, high data
transfer rates, fast seeking time, and last but not least, good 
reliability.
It is connected to the DAQ CPU through the built-in SCSI-2 interface.

\subsubsection{Trigger handshake}

The DISTO trigger system, described in detail elsewhere~\cite{distotrig},
is completely managed by the DAQ system. Two configurations
of trigger set-up were used throughout the experiment:
\begin{enumerate}
  \item multiparticle trigger for the $\vec{p} p \rightarrow p K^{+} \vec{Y}$ 
        and $\vec{p} p \rightarrow p p \phi$ measurements.  
  \item monitor trigger for $\vec{p} p$ elastic scattering reactions used to
        check and normalize the primary measurements. 
\end{enumerate}

The trigger is completely configurable by software by an independent 
program which produces two configuration files ready to be downloaded by
DAQ to the level-1 trigger CAMAC modules.
The number of beam spills for each trigger configuration can be 
selected at run-time 
by an appropriate choice of two parameters included in the set-up file.
  

\vspace{2mm}
\begin{figure}
   \centering\epsfig{figure=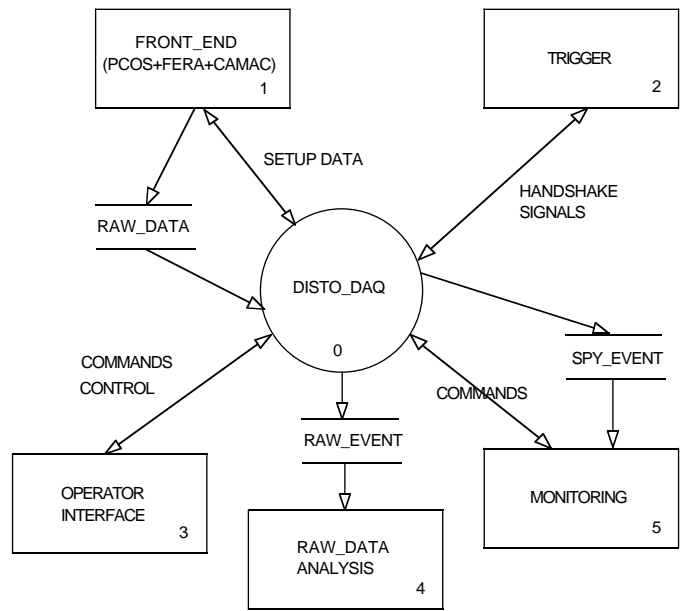,width=\linewidth}
\caption{\em The data acquisition context scheme}
\label{fig_contest}
\end{figure}

\subsection{Data Acquisition Software Architecture}

The DAQ software is organized as different tasks 
all running concurrently and cooperatively on different processors 
in the VME crate.
The run control, recording and acquisition tasks run under the 
operating system OS9 on the 68030 processor.
The PAW based monitoring task runs under UNIX in the SGI and DEC workstations 
and the 2nd level trigger task runs on a battery of four RISC 
CPU processors.

\begin{figure}
   \centering\epsfig{figure=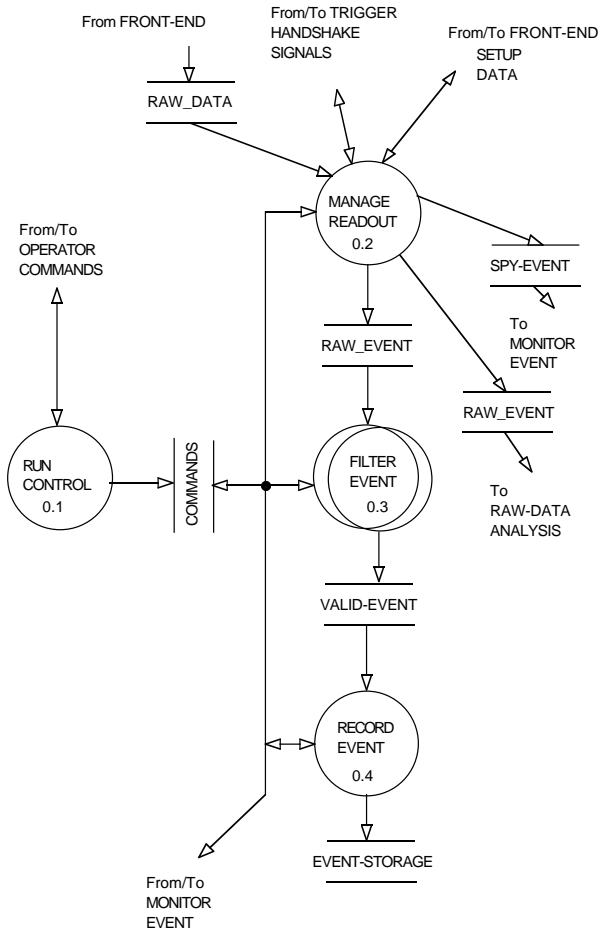,width=0.9\linewidth}
\caption{\em Overview of the data flow diagram and tasks.}
\label{fig_daq}
\end{figure}

In the context scheme of Fig.~\ref{fig_contest} the structure of the 
system is shown together with its environment.
DISTO\_DAQ includes all the tasks, with the exception 
of monitoring, which together with the front-end electronics, 
the trigger system, the operator interface and the raw data analysis 
make up its environment.
An overview of DISTO\_DAQ  is presented in Fig.~\ref{fig_daq}.

\vspace{2mm}

The MANAGE READOUT is the central part of the acquisition.
It receives RAW\_DATA from the front-end electronics 
and creates the buffers for the 2nd level trigger, for the 
monitoring and for the raw data analysis processor.
Moreover at the start of run it is able to initialize and check the dynamical 
hardware configuration and set all
parameters.
In the testing phase it can compute the FERA pedestals, 
the PCOS delays and execute front-end electronic tests.
During acquisition the task is able to compute 
the trigger type and include or inhibit the FERA pedestal subtraction 
and zero and overflow suppression as determined by the configuration file. 
The hardware configuration parameters and all other information 
to set up the  system (delays, thresholds, mapping logic addresses, 
pedestals, trigger type, etc.) are supported by disk files, which 
one can edit and download at the start of run.

\begin{figure}
   \centering\epsfig{figure=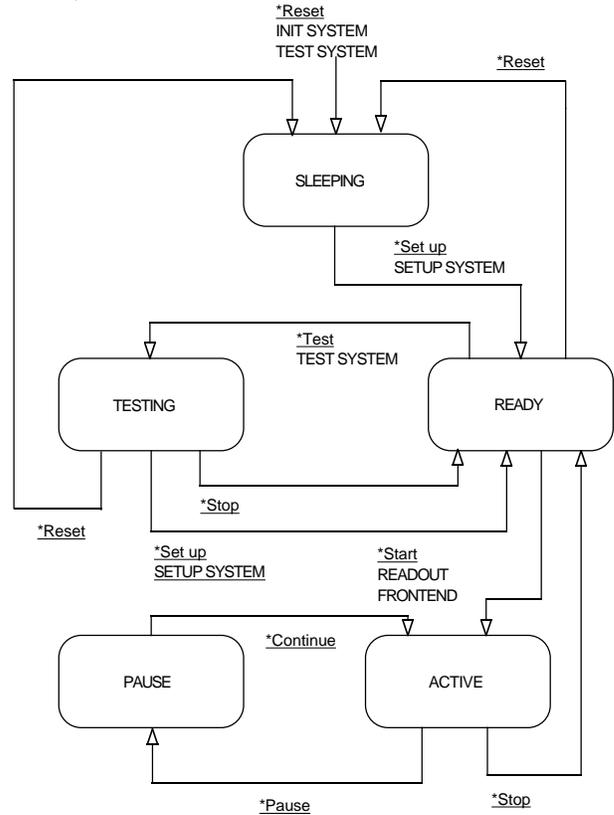,width=0.9\linewidth}
\caption{\em The data acquisition state machine}
\label{fig_state_machine}
\end{figure}

Operation of the data acquisition system is managed 
by the Run-Control sub-system, consisting of a concurrent process 
and one operator interface.
Run-Control manages all other DAQ sub-system 
tasks following a common state machine model with state transition 
corresponding to operator commands to start, stop, pause or end 
a run (period of data taking).
A simplified state machine is displayed in Fig.~\ref{fig_state_machine}.
All the synchronizations between various tasks 
are performed by the use of shared dual port memories on the 
VME-bus.

%
%
\subsubsection{Event Building}

The event building is carried out by both hardware and software systems.  
(Fig.~\ref{fig_daq_chain} and Fig.~\ref{fig_daq})
First, the event components coming from the 
detector readout electronics are collected into two cascaded triple port 
ECL memories.
In particular, the trigger signal makes the nine PCOS crates start to send 
their data in parallel to nine FIFO modules that subsequently transfers data 
sequentially on the ECL bus connected by the  multiplexer  to the ECL memory.  
Next, a signal switches the multiplexer input, linking it to the ECL bus 
coming from the FERA system.  The FERA starts to send its data on the ECL bus 
to the memory, queuing them after the PCOS data.  The last FERA data is the 
MARKER event separator, necessary because all events produced during the 
Saturne spill are queued in the same 
memory.  At the end of this time a signal communicates to the acquisition task 
that the data of the burst are ready in the memory and the software phase 
starts.

  The first step taken by the acquisition task is to build the event buffer, 
containing all of the events produced during the burst.  Inserted at the top 
of the buffer is the header information:  buffer length, run number, error 
types, trigger information and so on.  Following the header comes scaler and 
register data.  Last, PCOS and FERA data for each event, stored in the triple 
port ECL memory, are inserted.  The complete buffer is delivered to the first 
free RISC processor running the second trigger level which produces a buffer 
optionally containing only filtered events ready to be saved  on the tape.  
The same buffer is delivered to the SGI and DEC workstations for the 
raw data analysis processing.  The buffer 
communication between the tasks is synchronized using software flags.

%
%
\subsubsection{Monitoring system}

The data taking was monitored using two independent systems.
In the SGI workstation a sampling of the data is made to monitor the detectors
performances and give an event display.
The DEC workstation monitors the Cerenkov detector and gives another on-line
event display.
These tasks are made on both workstations under the PAW system~\cite{paw}
developed at CERN.

 %
%
\subsubsection{Run control system}
  
  The acquisition system task architecture is based on a states structure 
(Fig. ~\ref{fig_state_machine}).  All the tasks are organized around the same 
algorithm state machine (ASM).  Moreover the tasks are structured and 
organized around a hierarchical configuration.  The 'Run Control' (master 
task) controls the ASM, receives the operator commands by  keyboard and 
delivers them to all other tasks (slaves).  The tasks execute the operations 
expressed by the command and change the present state; at the end, all the 
tasks (also the run control) are always in the same state indicated by the 
ASM.  The ASM is composed of 5 states:
\begin{description}
    \item[SLEEPING]	This is an idle state; at the start all the tasks 
                        enter in this state.
    \item[READY]	In this state the system is ready for whatever 
                        operation. \\
            		It is possible to return to SLEEPING or to go to 
                        TESTING state to execute tests on the 
                        system hardware components or to go to the ACTIVE 
                        state to start data acquisition.
    \item[ACTIVE]	This is the data acquisition state.

    \item[TESTING]	In this state the system can do several tests on all 
                        the system hardware components.
    \item[PAUSE]	It is the state to suspend (temporarily) the data 
                        acquisition.
\end{description}
     
  The 'Run Control' also has control of the  general system state, continually
monitoring the real task situation, showing possible abnormal conditions.    
The structure used to synchronize all the operations is a shared memory 
housed in the VME bus.
%
%
\section{Performances}

The DISTO Data Acquisition System architecture, in spite of its apparent 
complexity, is simple and  compact.   It has proven to be extremely efficient 
and fast due to the choice of putting some key components in hardware (such as 
digitization and buffering), therefore only a small part of the software has 
to manage the real-time aspects of the data flow.

At present the system is able to collect data at 
a rate greater than 10000 ev/burst with the FERA zero and overflow 
suppression (the PCOS is intrinsically zero suppressed).  The event size is 
240 bytes for the multiparticle trigger and 160 bytes for the $p p$ elastic 
scattering monitor trigger.  The trigger rate is presently reduced to $\approx 
3500$ ev/burst for multiparticle trigger and $\approx 1700$ ev/burst for
monitor trigger, by a sharp trigger topology request and beam intensity 
modulation, in order to minimize the off-line correction due to read-out 
dead-time.  In production running of the experiment the mean read-out time is 
$\approx 19~\mu\,s$ and, consequently, the dead-time is 15\% for multiparticle 
trigger and 10\% for monitor trigger.
These rates were obtained for an instantaneous luminosity of 
$10^{31}$ cm$^{-2}$ s$^{-1}$.
\section{Acknowledgments}

  We wish to thank S. Gallian, and G. Maniscalco for their 
contribution to the DAQ and detectors electronic.  
  We are moreover greatly grateful to G. Abbrugiati, G. Giraudo, and M. 
Mucchi, for their effort in the design and construction of scintillating 
fiber detectors.
  We are particularly indebted to N. Dibiase for the essential 
support since the beginning of the experiment.
  The continous help of the whole Saturne staff was essential for the
success of this work.


\begin{thebibliography}{99}

  \bibitem{arvieux1} J. Arvieux et al.,
                        \textit{Proposal 213 - Saturne} (1991).

  \bibitem{abegg}    R. Abegg et al.,
                        \textit{Addendum 281 - Saturne} (1993).

  \bibitem{vigdor}   S. E. Vigdor,
                     \textit{Flavour and Spin in Hadronic and
                     Electromagnetic Interaction} eds. F. Balestra, 
                     R. Bertini, and  R. Garfagnini (1993) p.317.

  \bibitem{arvieux2} J.Arvieux et al.,
                     \textit{Int. Symposium on Polarization Phenomena in
                     Nuclear Physics} eds. E.J. Stephenson and S.E. Vigdor,
                     AIP Conf. Proceedings No. 339 (1994) p. 476.

  \bibitem{bertini}  R. Bertini, \textit{Nucl. Phys.} \underline{A585}
                     (1995) 265c.

  \bibitem{maggiora} A. Maggiora, \textit{Nucl. Phys. News} Vol.5 No.4
                        (1995) 23.

  \bibitem{eltec} ELTEC Elektronik GmbH, 
                  \textit{Galileo Galilei Stra{\ss}e 11,
                         D-55129 Mainz, GERMANY}.

  \bibitem{ces} CES S.A., \textit{70, Rue du Pont Butin, PO Box 107,
                         CH-1213 Petit Lancy 1, SWITZERLAND}.

  \bibitem{paw} R. Brun et al., 
                      \textit{PAW -- Physics Analysis Workstation}, \\ 
                      CERN program library long writeup Q121.

  \bibitem{lecroy} LeCroy Corporate, \textit{700 Chestnut Ridge Road, 
                         Chestnut Ridge, NY 10977-6499}.

  \bibitem{nanom} Nanometrics Systems Inc.,
                        \textit{Oak Park, Illinois}.

  \bibitem{caen} CAEN SpA,
                        \textit{Viareggio, Italy}.

  \bibitem{distotrig} F. Balestra et al.,
                        \textit{this proceeding}.

\end{thebibliography}
\end{document}